# Nanodomain Structure and Function in HTSC

## J. C. Phillips[1] and J. Jung[2]

1. Bell Labs., Lucent Tech. (Retired),  Murray Hill, N. J. 07974-0636

2. Dept. Physics, Univ. of Alberta, Edmonton, AB  T6G 2J1, Canada

The causes of high-temperature superconductivity (HTSC) are still mysterious, although more than 50,000 experiments have studied this subject.  The most severe test of any microscopic theory is generally considered to be its ability to *predict* the results of future experiments.  Here we examine recent (99-01) studies of BSCCO films by STM, which have revealed  nanodomain structure on a scale of 3 nm which is closely correlated with both superconductive gaps and pseudogaps.  This structure and these correlations were *predicted* as part of a discrete filamentary model of HTSC in 90.  The nanodomain diameter of 3 nm was identified in experiments on YBCO in 96.  While none of the experiments can directly establish causes, in the *predictive* theoretical model it was proposed that the underlying forces generating the nanostructure are ferroelastic.  It was also *predicted* that the strong correlations of the superconductive gap and pseudogap electronic structure with nanostructure are the result of dopant self-organization.  Here we describe a new method of preparing boride alloys, and we *predict* that it may produce materials with $T_c \sim 150$ K or more.

## I.  INTRODUCTION

Almost all experiments on the properties of high temperature superconductors are not directly sensitive to nanostructure.  Thus it has been customary to interpret these experiments in the context of the effective medium approximation (EMA).  (Fermi liquid theory is the simplest metallic realization of the EMA.) There is a growing awareness among experimenters that *native spatial inhomogeneities* play an important part in the anomalous properties of HTSC, as stressed very early (Mueller *et al.* 1987; Phillips 1987).  The latter proposed a specific nanostructural model (Phillips 1990) for HTSC in which nanodomains play an essential role.  They break up the $CuO_2$ planes into metallic nanoislands and semiconductive nanodomain wall networks that contain a pseudogap (Phillips 1987).  Thus the planes as a whole exhibit a pseudogap for electrical conduction confined to a single plane. However, currents can flow at frequencies below the gap by passing through resonant tunneling dopant centers in the semiconductive planes (such as BaO) separating the cuprate planes from other metallic planes. Zigzag filamentary paths of this type have recently been used to explain many anomalies in neutron (Phillips 2001; Reichardt *et al.* 1989) and infrared (especially c-axis (Homes *et al.* 1995)) vibronic spectra (Phillips and Jung 2001), and are illustrated in Fig. 1.

## II.     ARPES AND SCANNING TUNNELING MICROSCOPY

Zigzag filamentary paths are curvilinear, as are the near-Fermi energy basis states that are pinned to them. These states are not eigenstates of either momentum $\mathbf{k}$ or position $\mathbf{R}$, and there is no probe that can identify them directly. One can project these states onto either the surface wave vector $\mathbf{k}_s$ in angle-resolved photoemission (ARPES) experiments (Loeser *et al.* 1997), or on to position coordinates $\mathbf{R}_s$ in STM experiments, and observe either angularly dependent $\mathbf{k}_s$ effects, or spatially dependent $\mathbf{R}_s$ effects, in a residual superconductive gap $\Delta$. In the filamentary model *none* of these measured values of $\Delta(\mathbf{k}_s)$ or $\Delta(\mathbf{R}_s)$ correspond to the constant $\Delta$ of the BCS theory, as one must average $\Delta$ over an entire filamentary path to obtain its value for a single path, and over pairs of paths to obtain effective Cooper pair path interactions. One must therefore handle microscopic data carefully in order to avoid falling into descriptive self-deception as to the "reality" of various "observed" gaps.

We now turn to the data (Hudson *et al.* 1999; Pan *et al.* 2000; Howard *et al.* 2001) on very carefully cleaved micaceous BSCCO obtained by scanning tunneling microscopy (STM). These data have been discussed only in terms of an effective medium model ("d waves") introduced to describe ARPES experiments. This model has also been proposed to describe a variety of other experiments that many observers already believed to be indecisive and uninformative, and in some cases quite possibly based on artifacts (Neils and Van Harlingen 2000). In any event, the d wave pairing model never pretended to be more than an EMA framework for describing structurally insensitive data, and it laid no claims to *explaining* the causes of HTSC. Now the STM data have shown that the d wave model overlooked all the nanostructural features that correlate well with electronic properties (Howard *et al.* 2001), such as gaps and pseudogaps. We have shown elsewhere (Phillips and Jung 2001) that the chemical trends in undoped and Zn-doped nanodomain features (Hudson *et al.* 1999; Pan *et al.* 2000) are incompatible with d waves, but are fully explained by ferroelastic nanodomains, and will repeat that discussion here for completeness.

The STM experiments (Hudson *et al.* 1999; Pan *et al.* 2000) observe the centers and angular variation of states pinned to a narrow energy region near the Fermi energy. There are two cases of such nanoscale zero-bias defect or impurity states: native X defects, probably associated with O vacancies, and extrinsic Zn impurities. According to the d-wave model, the radial extent of the nearly zero-bias states should depend on the nature or strength of the impurity (X or Zn) potential. On the other hand, the angular behavior should be characteristic of the host d-wave pairing itself and should be independent of the nature of the impurity. The nanodomain model predicts just the opposite: the radial extent of the zero-bias states depends only on the host composition-independent, ferroelastically fixed nanodomain size of 3 nm, and is nearly independent of the nature of the impurity.

The angular behavior is determined by whether or not the impurity is located within the domain, or at a domain wall or wall intersection. The X impurities are native and are most likely to be found in the domains. The natural site for the Zn in the $CuO_2$ planes is at the corners of the nanodomains, as the nanodomain walls function as relievers of interplanar $CuO_2$-BaO ferroelastic misfit. Generally the bonding distortions at the domain walls must already be large, and so the strain energy will be minimized if the Zn occupies a more distorted wall site rather than an undistorted domain interior site. Moreover, if the Zn does even better, and occupies a corner site, it is likely to quench the local pseudogap associated with the neighborhood of that site, converting the four associated nanodomains from coherent and filamentary regions, to Fermi liquid (effectvely overdoped) non-superconductive regions. Note that there are $\sim n^2$ unit cells/nanodomain, and that a corner Zn will affect four nanodomains. With nanodomain diameters of 3 nm and the planar lattice constant a $\sim 0.5$ nm, there are n $\sim 6$ unit cells in a nanodomain diameter. This means that $\sim 1/(4n^2) \sim 1\%$ Zn can quench both superconductivity and filamentary vibronic anomalies, in agreement with STM and infrared experiments (Phillips and Jung 2001). Of course, d wave pairing makes no quantitative predictions at all concerning the critical Zn concentration, because it is merely descriptive and is not a comprehensive theory.

(Pan *et al.* 2000) note that in addition to the zero-bias bright spot associated with each Zn atom, one can discern narrow tetragonal spokes emanating up to 3 nm from a spot. These spokes are not observed in connection with impurity or defect X states that produce zero-bias bright spots in undoped BSCCO. They interpret these spokes as evidence for d-wave superconductivity, which is supposed to be a characteristic feature of the host, and so should be observed in connection with either Zn states or X states. We prefer to interpret these spokes as representing both the anisotropy of the pseudogap which is localized in the nanodomain walls, as well as the anisotropy of the Zn resonant charge density that is centered at the intersections of these walls. Then the difference between the STS images for the X states and the Zn states is understood by the Zn site preference for nanodomain corners.

In passing, it is worth noting that in the published images the spokes are quite narrow, and one would assume that the original images were even narrower. Their angular variation (on a logarithmic scale!) appears to be much too rapid to be described by a merely quadratic d function, such as $F(x,y) = (x^2 - y^2)$, and a function such as $F^{2n}$, with n >> 1, or even an exponential, would appear to give a much better fit. Where such rapidly varying angular terms would originate in a continuum model is difficult to imagine. However, in the nanodomain model the ratio (a/d) of nanodomain wall thickness a to nanodomain diameter d is expected to be ~ 1/6. This, together with the expected exponential attenuation of the Zn dopant zero-bias charge density outside the domain walls, would readily explain the observed narrowness of the tetragonal spokes.

Ordinarily to distinguish between two competing theoretical models one asks for a single decisive experiment. Here we have *three* decisive experiments, all of which strongly support the ancient 1987, 1990 filamentary nanodomain model, and strongly disagree with the more recent d wave pairing model. To our knowledge the ancient filamentary nanodomain model is the only theoretical model proposed so far that has been able to achieve such predictive success.

## III. ORIGIN OF NANODOMAINS

The easiest way to understand where nanodomain structural features come from, and what factors determine their dimensions, is by analogy with the very well-established theory of misfit dislocations, which is reviewed here for the reader's convenience. Mathematical studies of heterointerfaces between thick films and crystalline substrates with one atom per unit cell and with different lattice constants showed long ago (Van der Merwe, 1963) very large effects, even for small misfits. As the lattice constant misfit stress increases, because the harmonic strain misfit energy grows *quadratically* with distance, at some point misfit dislocations occur which lower the elastic energy of the system relative to uniform elastic strain. (Note also that electronic phase energy differences also grow only *linearly* with the filling factor, so that the elastic energies, while small at short distances, must always overwhelm the electronic ones at nanometer length scales. This point is generally overlooked in theoretical discussions (Zaanen 1998) of charge-spin separation that ignore elastic interactions.) In the semiconductor case, the misfit dislocation is often pictured merely as a half-plane of extra atoms, but this is not the only possibility; elastic misfit can be relieved equally well by any second phase with a different effective lattice constant. In particular, if the energy differences of the two phases are small enough, the critical layer thickness at which a lower total energy is achieved in the mixed (phase-separated) state relative to a homogeneous phase can be reduced to that of a single atomic layer.

These misfit ideas are well-known to crystal chemists (Goodenough and Manthiram 1990; Wilson 1998), and the most commonly discussed example of misfit is the cubic perovskites, such as $ABO_3$. Because there is only one cubic lattice constant, if the structure remains cubic (because this is favored by long-range ionic forces), then the A-O and B-O bond lengths will be strained. These old ideas can be generalized to more complex pseudoperovskite layered structures by constraint theory (Phillips 1987). Some of this ferroelastic strain energy can be relieved, in some cases, by special magnetic or electric structures, for example, ferroelectricity in $BaTiO_3$ and other perovskites (Nambu and Sagala 1994; Aleksandrov 1995; Ivanov 1998). In the layered cuprates, with strong intraplanar interactions and weaker interplanar ones, some strain energy is required to match the planar lattice constants of the cuprate layers and the semiconductive layers (described as the chess-board and rock-salt layers by (Wilson 1998)); this strain energy can be reduced by uniaxial strains which leave the unit cell volume little changed, and this is the general explanation behind the origin of the split apical oxygen sites. With a 12% axial apical O splitting, at constant volume a planar lattice constant misfit of order 6% is implied; based on bond radius concepts that are probably inaccurate, an even larger value of 13% has been estimated (Goodenough and Manthiram 1990). These estimates predict nanodomain diameters of order 2- 4 nm.

The first reliable estimate of nanodomain dimensions (~ 2.5 nm) was obtained in TEM studies which produced a detailed ferroelastic structural model for YBCO (Etheridge 1996). [She commented that these domains would be coupled through domain walls by the proximity effect. Thus one could see a superconductive gap $\Delta'$ in the domain walls that is smaller than the gap $\Delta$ seen in the domains. This indeed appears to have been observed, $\Delta' \sim \Delta/2$ in the STM data on the Bi surface layer (Howard *et al.* 2001).] A similar estimate (3-4 nm) was obtained in critical current density studies in YBCO that identified the crossover where the coherence length equals the nanodomain size (Darhmaoui and Jung 1996).

A number of ferroelastic structural anomalies occur at or very close to $T_c$, and by now it is clear that it is extremely unlikely that all these coincidences are accidental. Moreover, such phase instabilities have been observed even in ultrathin film semiconductors with the very simple cubic zincblende structure and no transition metals or oxygen. For instance, in CdTe/ZnTe ultrathin multi-quantum wells (2-4 atomic planes), a dynamical Jahn-Teller effect has been observed in the width of the exciton Urbach tail, which is minimized at 45K, below which tail-broadening reconstruction occurs. The authors related the reconstruction to interfacial strain which would have produced misfit dislocations for thicker layers (Yoshimura, Nakata, Ohyama, and Otsuka 1993).

Summarizing, we can say that the evidence that nanodomains arise primarily from interlayer ferroelastic misfit, and not from incidental "charge-spin separation", or what have you, is firm. The pseudogap in the domain walls is simply a one-dimensional Peierls instability related to density waves, either charge or spin, or both, that generates a shifted effective lattice constant that reduces interlayer misfit. But what is the function of these nanodomains? Are they the only factor that is necessary for HTSC or CMR (colossal magnetoresistance)?

## IV. DOPANT SELF-ORGANIZATION

The reader has doubtless already guessed that there is a second factor that is necessary to produce electronic anomalies such as HTSC and CMR. This second factor is the crucial one that is always omitted in order to simplify calculations in continuum models based on the EMA, such as the rigid band model. It is the positions of the dopants. Although the dopants are not subject to long-range order, their disorder is not random. Instead, because of their high mobility, native dopants (such as oxygen vacancies) adopt configurations during high-temperature annealing that maximize the electronic conductivity. (Such maximization minimizes the free energy by optimally screening internal fluctuating ionic fields.) In the absence of the glassy nanodomain grid, the positions of the dopants would not be of much importance. However, in the presence of such a grid the conductivity can be greatly increased (essentially from insulating to metallic) by placing the dopants in an especially efficient topological configuration such that there are exactly two (in/out, or source/drain) electrically active resonant tunneling centers per nanodomain (see Fig. 1). This makes possible a fundamental reordering of energy levels near the Fermi energy (within the pseudogap) to form a narrow, dopant centered high-mobility impurity band.

The special configuration of dopants is said to be *self-organized*. This phrase, which seems to have originated with Prigogine in his early efforts (~ 1950) to answer Heisenberg's famous 1945 question, What Is Life?, has been widely discussed recently in the context of self-organized criticality (pinned charge density waves, sand piles, avalanches and forest fires) (Turcotte 1999). The kind of filamentary self-organization discussed here is topologically much more interesting, much more active, and comes much closer to answering Heisenberg's question. In fact, filamentary self-organization is mathematically very close to evolutionary biology, as can be seen by studying scaling exponents for metabolism vs. mass (Phillips 2000). In any event, the filamentary dopant centered high-mobility impurity band explains normal-state transport anomalies (Stormer et al. 1988) which give resistivities and Hall numbers linear in T, and infrared conductivities nearly linear in $\omega$ (Dodge *et al.* 2000).

## V. ARE STRIPES SELF-ORGANIZED?

Some authors (Emery *et al.* 1997) have taken the view that the stripe topology that has been found in $La_{2-x}Sr_xCuO_4$ near $x = 1/8$ is "somehow" a paradigm for the kind of spatial inhomogeneities that generate HTSC and CMR. To see what "somehow" really means, let us first discuss the phase diagram for $La_{2-x}Sr_xCuO_4$. The stripe (1/8) phase (Tranquada 1998) occupies the range $0.12 < x < 0.15$, and it is adjacent to the filamentary phase, which fills the range $0.15 < x < 0.21$. The stripe phase consists of alternating orthorhombic crystalline strips, one antiferromagnetic, and one a Fermi liquid; neither is superconductive. The filamentary phase consists of a partially disordered mixture of orthorhombic nanodomains of both planar polarizations (Haskel *et al.* 1996). The strength of the minority nanodomain ferroelastic polarization increases from zero at $x = 0.15$ to equal that of the majority at $x = 0.21$, where a transition from the (directed) filamentary HTSC to the non-superconductive Fermi liquid takes place. Residual superconductivity is observed for $x < 0.15$, or $x > 0.21$, only because the Sr dopants in $La_{2-x}Sr_xCuO_4$ ($T_c \sim 40K$) are much less mobile than the oxygen dopants in HTSC with $T_c \sim 90K$. Thus these samples are not fully equilibrated and homogenized, and the Sr dopant configurations are not completely self-organized.

To convert stripes into filaments, the following steps must take place. (1) The long-range orthorhombic strip ordering must become medium range ($\sim 3$ nm) and glassy (partially disordered), so that it is no longer accessible to conventional diffraction experiments. (2) The antiferromagnetic strips must turn into a planar network of nanodomain walls. (3) The dopants must self-organize to produce coherent filaments.

This will be possible only in a range where the dopant density that has not precipitated in the domain walls is such that filamentary percolation can occur (Phillips 1999). While nature itself seems to have had little difficulty in executing these steps, it is unlikely that the picture of stripes by itself would have led any mere human to visualize this particular "somehow", especially as the proposed mechanism for HTSC in the stripe model involves (Emery *et al.* 1997) virtual exchange of *magnons*, not phonons.

A characteristic topological feature of the filamentary model is that the filaments must contain very few small (electronically diamagnetic) closed loops. This point has been demonstrated in numerical simulations of self-organized intermediate phases in network glasses (Thorpe *et al.* 2000). The fact that the elastic and HTSC filamentary networks behave so similarly in topological terms is not accidental. For central force models it is believed that the percolative threshold and scaling exponent are the same for shear waves and the electrical conductivity (Plischke and Joos, 1998). The local orthorhombicity is characterized by tilting of $CuO_6$ octahedra, well-described by central forces, as rigid units.

Filamentary self-organization is highly non-trivial; indeed, without it, one would have only conventional site percolation, for which there is only one metal-insulator phase transition, no intermediate phase, and neither HTSC nor CMR. In fact, the rigid stripe structure, which lacks both flexible medium-range disorder, and dopant self-organization, is an excellent example of a system that constitutes a phase that is neither insulating nor a Fermi liquid, but nevertheless, does not exhibit either HTSC or CMR. In other words, it is not enough merely to form a nanoscale phase-separated phase, as this can be done in many ways, while only one way produces HTSC. It is striking that in all the discussions (some quite lengthy (Emery *et al.* 1997), involving more than 200 references), all based on the EMA, of how the stripe phase might "somehow" become a HTSC, we have found no mention of ferroelasticity, nanodomains, glassy disorder, phonons, or dopant self-organization.

## VI.   HOW TO RAISE $T_C$ IN THE BORIDES

The discovery of superconductivity in $MgB_2$ with $T_c$ = 39 K has surprised many people, as the transition temperature is far higher than was ever attained in ternary borides (Matthias *et al.* 1977). So far, however, many of the properties seem to resemble those of old superconductors like $Nb_3Sn$ (Finnemore *et al* 2001). A theoretical model, based on LAPW band structure calculations, has led to the conclusion (Kortus *et al* 2001) that the compound is strongly ionic. All of the Mg valence electrons are donated to the B valence band, so that $T_c$ is high because the band width W is large, as it might be in metallic hydrogen. This is a picture that is chemically surprising (most conventional superconductors have much stronger electron-phonon interactions with heavy elements with polarizable soft cores, so that chemical trends in the V factor ($\sim$ W) dominate those in N(0) $\sim$ Z/W in the coupling factor N(0)V), and the authors admit that it is exotic. We distrust this picture, because $MgB_2$ has a network structure, and traditionally the LAPW method (designed to treat close-packed metals) has been unreliable in estimating charge transfer in open network structures such as semiconductors.

In data (Finnemore *et al* 2001) plotting resistivity as a function of T (0-300K) in various magnetic fields (0-9T), there is one feature which is *not* seen in old superconductors like $Nb_3Sn$. This is a large magnetoresistance, extending from $T_c$ = 39 K up to $\sim$ 150K. Such large magnetoresistance can be the result of domain wall scattering of either spins or magnetically distorted filaments, provided the domain walls are narrow and reasonably closely spaced (Tatara *et al.* 1999). In other words, there may be incipient nanodomain formation, and this would enhance $T_c$. If this is so, it raises the possibility that very high $T_c$ 's might be attainable by exploiting the recipe used to raise $T_c$ from 1 K to 90K in $Na_xWO_3$ (Reich and Tsabba 1999). This recipe involves a lot of high-temperature synthetic magic. In the present case one should mix Mg, B and Cd or Hg in an evacuated glass tube and anneal at high T for long times, followed by very slow cooling. This could produce a substrate of $MgB_2$ with epitaxial surface islands (Levi *et al.* 2000) of $(Cd$ or $Hg)_xB_2$.

The key point now is that, because of the size differences between Mg and Cd or Hg, in order for $(Cd$ or $Hg)_xB_2$ to grow pseudoepitaxially on $MgB_2$, it is likely that x < 1, in other words, the islands are likely to be metastable defect phases. According to the filamentary theory (Phillips, 1987, 1990) this is good, not bad, because a defect Stormer band can be formed that will pin the Fermi energy and have very large e-p interactions. Moreover, the defects (which can be either the cations or their vacancies) can self-organize to form filaments. The anomalous magnetoresistance $MgB_2$ data discussed above suggest that this may already have happened to some extent there, but the possibility exists of a much larger effect, and much higher $T_c$'s, in metastable (Cd or $Hg)_xB_2$.

## VII.    CONCLUSIONS

In HTSC the basic planar length scale (coherence length) is set by nanodomain diameters (Darhmaoui and Jung 1996), and the presence of glassy planar nanodomains completely alters all the physical properties. The nanodomains, and only the nanodomains, make possible a discrete intermediate phase (the inaptly and ineptly named "non-Fermi liquid") that exhibits HTSC. Thus the study of HTSC at the microscopic level becomes the study of nanodomain structure and function. This creates a range of theoretical problems that in the past have been avoided by making the EMA. Unfortunately, if the EMA were valid, there would be no HTSC. Thus these problems must be addressed, not avoided, and the question is, how best to do this.

Whenever a large-area study is made of glassy domains, statistical problems arise, even in very simple cases such as giant domains observed by transmission electron microscopy of ultrapure chalcogenide alloy glass films (Chen *et al.*1984). (This work was featured on the cover of *Physics Today* (Phillips 1982)). These problems have frustrated many theorists for many years. For example, one eminent theorist has worked in this field for nearly 50 years, and has produced a few results on electronic transport, which have unfortunately been disproved by recent experiments (Itoh *et al.* 1999). An even more eminent theorist has discussed this subject for more than 40 years, without producing any results that could be applied to electronic transport. In HTSC the problems are much more difficult, as the electronic structure varies drastically from nanodomain center to nanodomain wall, and the data can be presented in quite different ways. Moreover, essential aspects of the filamentary physics (the c-axis cross-links) are inaccessible to observation by STM. Thus a complete picture can be obtained only by linking the STM data to a theory that provides a consistent platform for discussing a wide range of experimental data. It is unlikely that such a platform will emerge from "numerous discussions" with theorists with no experience in the physics of disorder (Phillips and Thorpe 2001).

Here we have argued that at present there is only one such platform, the filamentary nanodomain model. This platform is part of a general approach to network connectivity transitions that has proved to be very, very successful in discussing the simplest of such transitions, the paradigmatic elastic stiffness transition in network glasses. The filamentary nanodomain model, in contrast to all alternative EMA theories (such as d-wave pairing and stripes) *predicted* from a very early stage (Phillips, 1987, 1990) that nanodomains and nanoscale phase separation are essential parts of a consistent theory of HTSC. The *quantitative* features reported in recent STM studies (Hudson *et al.* 1999; Pan *et al.* 2000) were interpreted there as supporting d wave pairing, but we have shown that those data actually exclude d wave pairing. (Howard *et al.* 2001), looking at very similar data, but differently presented, reached the same conclusion, but were unable to offer an alternative theory. We have shown here that all of the apparently contrasting effects discussed in both studies were already predicted by the filamentary model long ago.

## APPENDIX

Because of its radically different character, HTSC from the beginning has been drastically different from most other fields of condensed matter science. It has had a wild "Woodstock" or uncritical "Anything Goes" style, both in theory and in experiment. Now 15 years after its discovery, there is a firm and growing base of excellent experimental data, and it is appropriate to ask why the standards for theoretical interpretations should not also be improved. This is why there has been so much emphasis here on the many successful nanostructural *predictions* of the filamentary nanodomain model.

Superconductivity seems, from its outset, to have had a serendipitous quality, so far as experiment is concerned: certainly Kamerling Onnes was not looking for SC when he (or his student) discovered it. This has led many experimentalists, and even quite a few theorists, to suppose that it is possible to get the correct theory of HTSC *by accident.* In our view this is nonsense: so far as we know, there is no example in all of theoretical physics, involving SC or any other phenomenon, of such an accident. The BCS theory is correct, and it was not an accident.

In fact, BCS (Bardeen *et al.* 1957) were careful to emphasize that their theory was correct because of its *predictive* aspects. In the relation $E_g = nkT_c$, their theory *predicted* n = 3.5, in excellent agreement with the data then available, and in even better agreement with later data. The theory *predicts* the ultrasonic and magnetic resonance relaxation critical behavior. Similarly here we have shown many points of agreement between the *predictions* of the filamentary nanodomain model (Phillips 1987, 1990; Darhmaoui and Jung 1996) and recent (99-01) STM experiments. That these successes should be accidental is no more likely than in any other case of successful theories.

# Figure Caption

Fig. 1.   The basic idea of the filamentary paths in the quantum percolative model (Phillips, 1987, 1990, and many others) for YBCO.   The positions of the **I**nsulating **N**anodomain **W**alls in the $CuO_2$ layers are indicated, together with the **R**esonating **T**unneling **C**enters in the semiconductive layer, and oxygen vacancies in the $CuO_{1-x}$ chains. Giant e-p interactions are associated with the **R.T.C.**, where the interactions with LO c-axis phonons are especially large.   The **I.N.W.** are perovskite-specific.   The *sharp bends* in the filamentary paths are responsible for the broken symmetry that admixes ab planar background currents with c-axis LO phonons.

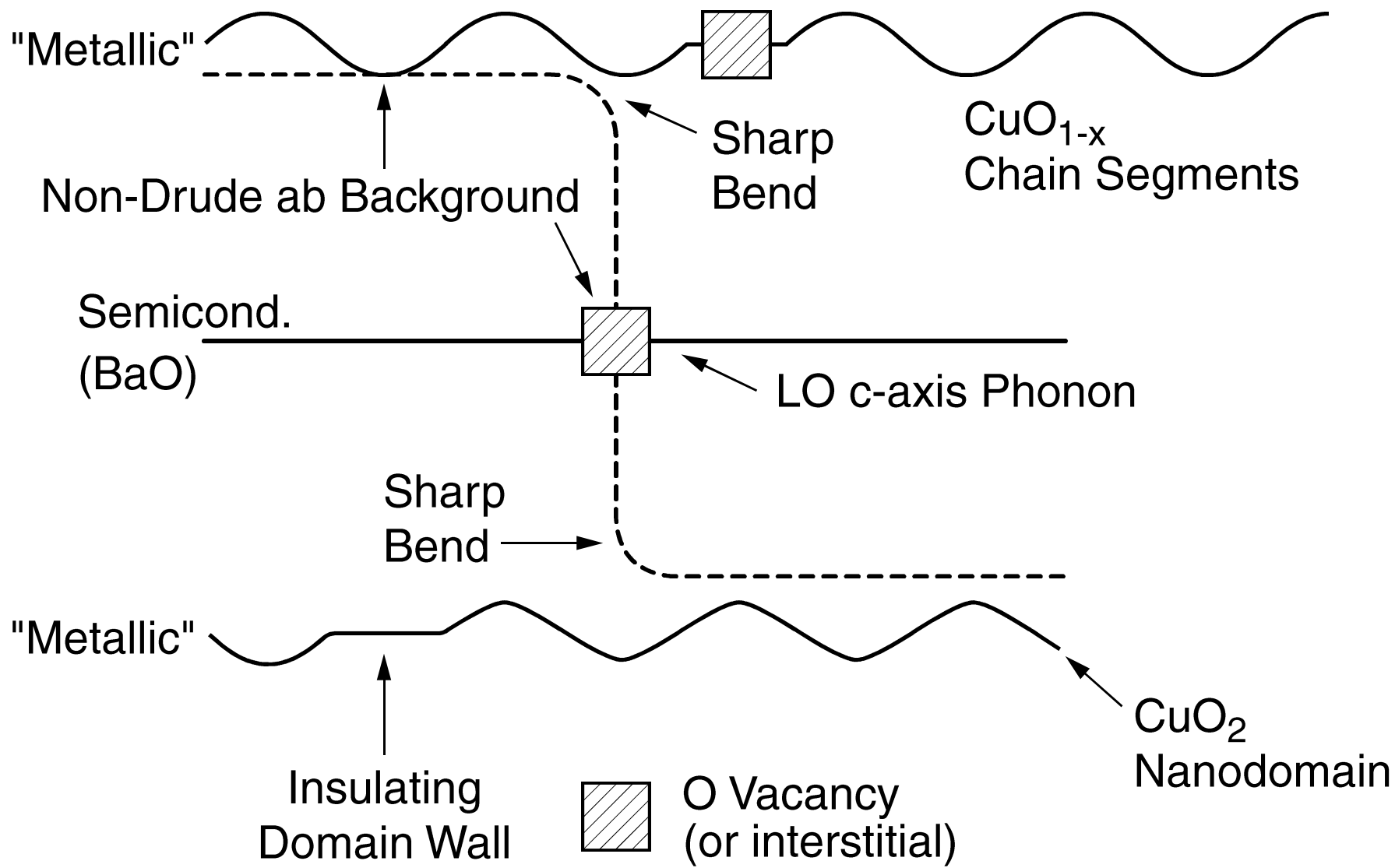

"Metallic"

Non-Drude ab Background

Sharp Bend

$CuO_{1-x}$ Chain Segments

Semicond. (BaO)

LO c-axis Phonon

Sharp Bend

"Metallic"

Insulating Domain Wall

O Vacancy (or interstitial)

$CuO_2$ Nanodomain